\documentclass[useAMS,usenatbib]{mn2e}
\usepackage{psfig}
\usepackage[a4paper]{hyperref}
\usepackage{amssymb}
\usepackage{aas_macros}

\newcommand{\rmsub}[2]{#1_{\rm #2}} 

\title[Transit search algorithm]{A fast hybrid algorithm for exoplanetary transit searches}
\author[A. Collier Cameron et al]
{
A. Collier Cameron$^{1}$\thanks{E-mail:acc4@st-and.ac.uk},
D. Pollacco$^{2}$,
R. A. Street$^{2}$,
T.A. Lister$^{1,5}$,
R.G. West $^{3}$,
\newauthor
D.M. Wilson $^{5}$,
F. Pont $^{10}$,
D.J. Christian$^{2}$,
W.I. Clarkson$^{4}$,
B. Enoch$^{4}$,
A. Evans$^{5}$,
\newauthor
A. Fitzsimmons$^{2}$,
C.A. Haswell$^{4}$,
C. Hellier$^{5}$,
S.T. Hodgkin$^{6}$,
K. Horne$^{1}$,
\newauthor
J. Irwin$^{6}$,
S.R. Kane$^{7}$,
F.P. Keenan$^{2}$,
A.J. Norton$^{4}$,
N.R. Parley$^{4}$,
J. Osborne$^{3}$,
\newauthor
R. Ryans$^{2}$,
I. Skillen$^{8}$ 
and
P.J. Wheatley $^{9}$
\\
$^{1}$School of Physics and Astronomy, University of St Andrews, North Haugh, St Andrews, Fife KY16 9SS, UK.\\
$^{2}$Astrophysics Research Centre, Main Physics Building, School of Mathematics \&\ Physics, Queen's University, University Road, Belfast, BT7 1NN, UK.\\
$^{3}$Department of Physics and Astronomy, University of Leicester, Leicester, LE1 7RH, UK.\\
$^{4}$ Department of Physics and Astronomy, The Open University, Milton Keynes, MK7 6AA, UK.\\
$^{5}$Astrophysics Group, School of Physical and Geographical Sciences, Lennard-Jones Laboratories, Keele University, Staffordshire, ST5 5BG.\\
$^{6}$Institute of Astronomy, University of Cambridge, Madingley Road, Cambridge, CB3 0HA, UK.\\
$^{7}$Department of Astronomy, University of Florida, 211 Bryant Space Science Center, Gainesville, FL 32611-2055, USA.\\
$^{8}$Isaac Newton Group of Telescopes, Apartado de correos 321, E-38700 Santa Cruz de la Palma, Tenerife, Spain. \\
$^{9}$Department of Physics, University of Warwick, Coventry CV4 7AL, UK.\\
$^{10}$Observatoire de Gen\`eve, 51 Ch. des Maillettes, 1290 Sauverny, Switzerland.\\
}
\begin{document}

\date{Accepted 0000 December 00. Received 0000 December 00; in original form 21 June 2006}

\pagerange{\pageref{firstpage}--\pageref{lastpage}} \pubyear{2006}

\maketitle

\label{firstpage}

\begin{abstract}
We present a fast and efficient hybrid algorithm for selecting exoplanetary candidates from 
wide-field transit surveys. Our method is based on the widely-used {\sc SysRem} and 
Box Least-Squares (BLS) algorithms. Patterns of systematic error that are common to
all stars on the frame are mapped and eliminated using the {\sc SysRem} algorithm. The
remaining systematic errors caused by spatially localised flat-fielding and other errors 
are quantified using a boxcar-smoothing method. We show that the dimensions of the
search-parameter space can be reduced greatly by carrying out an initial BLS
search on a coarse grid of reduced dimensions, followed by Newton-Raphson refinement 
of the transit parameters in the vicinity of the most significant solutions. We illustrate the 
method's operation by applying it to data from one field of the SuperWASP survey, comprising 2300 observations of 7840 stars brighter than $V=13.0$. 
We identify 11 likely transit candidates. We reject stars that exhibit significant ellipsoidal variations indicative of a stellar-mass companion. We use colours and proper motions from the 2MASS and USNO-B1.0 surveys to estimate the stellar
parameters and the companion radius. We find that two stars showing unambiguous
transit signals  pass all these tests, and so qualify for detailed high-resolution spectroscopic follow-up.
\end{abstract}


\section{Introduction}

Among the 194 extra-solar planets discovered in the last
decade, the ``hot Jupiters" currently present the greatest challenges to understanding
and the greatest observational rewards. Efforts to explain their origin have
transformed theories of planetary-system formation. Do these planets form via
gravitational instability in cold discs, or must they form by core accretion beyond
the ice boundary, where  ice mantles on dust grains allow rapid agglomeration of a
massive core of heavy elements? How rapidly  do they migrate inwards through a
massive protoplanetary disc, and what mechanism halts the migration?

The subset of these planets that transit their parent stars are of key importance in
addressing these questions, because they are the only planets whose radii and masses
can be determined directly. The first such discovery \citep{charbonneau2000}
confirmed the gas-giant nature of the hot Jupiters; indeed the inflated radius of HD
209458b continues to challenge our understanding of exoplanetary interior structure.
Subsequent discoveries have presented further surprises. The apparent correlation
between planet mass and minimum survivable orbital separation among the known
transiting planets offers important clues to the ``stopping mechanism" for inward
orbital migration of newly-formed giant planets in protoplanetary discs \citep{mazeh2005}. 
The high core mass of the recently-discovered Saturn-mass planet orbiting HD
149026 (Sato et al. 2005)\nocite{sato2005} presents difficulties for models of planet formation via
gravitational instability. {\em Spitzer} observations of the secondary eclipses of TReS-1b, HD 209458b 
and HD189733b have provided the first estimates of the temperatures in these
planets' cloud decks (\citealt{charbonneau2005}; \citealt{deming2005}; \citealt{deming2006}). 

To date, however, only ten such planets have been
discovered. Three of the ten have been found via targeted radial-velocity searches,
the other six through large-scale photometric surveys for transit events. The primary
goal of
the SuperWASP project (Pollacco et al. 2006)\nocite{pollacco2006} is to discover bright, new transiting exoplanets in
sufficiently large numbers that we can place studies of their mass-radius relation,
and the suspected relationship between minimum orbital distance and planet mass, on a
secure statistical footing. SuperWASP's  ``shallow-but-wide" approach to
transit hunting is designed to find planets that are not only sufficiently bright
($10 < V < 13.0$) for high-precision radial-velocity follow-up studies to be feasible
on telescopes of modest aperture, but also for detailed follow-up studies such as
transmission spectroscopy during transits, and Spitzer secondary-eclipse
observations. Only five of the ten presently-known transiting planets orbit stars
brighter than $V=14$ and so have such strong follow-up potential. 

Here we present a methodology used in the search for exoplanetary transit candidates 
 in data from the first year of SuperWASP operation. We employ the {\sc SysRem} algorithm of 
 \citet{tamuz2005} to identify and remove patterns of correlated systematic error from the 
 stellar light-curves. We present a refined version of the  Box Least-Squares (BLS) 
 algorithm of \citet{kovacs2002}, which permits a fast grid search and efficient refinement
 of the most promising solutions without binning the data. Using simulated transits 
 injected into real SuperWASP data we develop
a filtering strategy to optimise and quantify the recovery rate and false-alarm probability 
as functions of stellar magnitude and transit depth. We develop simple plausibility 
tests for transit candidates, using the transit duration and depth to estimate the mass 
of the parent star and the radius of the planet. We mine publicly-available catalogues to
obtain the $V-K$ colours, proper motions and other properties of the host stars. 
Combined with the transit durations and depths, these 
give improved physical parameters of each system and help us to  identify the most 
promising candidates for spectroscopic follow-up observations.

\section{Instrumentation and Observations}

The SuperWASP camera array, located at the Observatorio del Roque de los Muchachos
on La Palma, Canary Islands, consists of five 200mm f/1.8 Canon lenses each with an 
Andor CCD array of $2048^2$ 13.5$\mu$m pixels, giving a field of view 7.8 degrees 
square for each camera. By cycling through a sequence of 7 or 8 fields located within
4 hours of the meridian, at field centres separated by one hour
right ascension, SuperWASP monitored up to 8\%\ of the entire sky for
between 4 and 8 hours each night during the 2004 observing
season. The average interval between visits to each field
was 6 minutes. Each exposure was of 30s duration, and was
taken without filters.

Between 2004 May and September, the five SuperWASP cameras
secured light-curves of some $2\times 10^5$ stars brighter than $V=13$. 
The long-term precision of the data (determined from the RMS scatter of 
individual data points for non-variable stars) is 0.004 mag at
$V=9.5$, degrading to 0.01 mag at $V=12.3$. 
Our sampling rate and run duration guarantee that 4 or more transits 
should have been observed 
in 90\% of all systems with periods less than
5 days, and 100\%\ of all systems with periods less than 4 days, though some
incompleteness is expected at periods very close to  integer multiples of 1 day. 

\section{Data reduction}

The data were reduced using the SuperWASP pipeline. The pipeline carries out an
initial statistical analysis of the raw images, classifying them as bias frames, flat fields, 
dark frames or object frames. The bias frames are combined using optimally weighted 
averaging with outlier rejection. Automated sequences of flat fields secured at dawn 
and dusk are corrected for sky illumination gradients and combined using an optimal 
algorithm that maps and corrects for the pattern introduced by the finite opening and  
closing time of the iris shutter on each camera. 

Science frames are bias-subtracted, 
corrected for shutter travel time and corrected for pixel-to-pixel sensitivity variations
and vignetting using the flat-field exposures. Flat fields from different nights are combined
using a algorithm in which the weights of older flat field frames decay on a timescale of 
14 days. A catalogue of objects on each frame is constructed using {\sc extractor}, 
the Starlink implementation of the {\sc SExtractor} \citep{bertin1996} source detection software.
An automated field recognition algorithm identifies the objects on the frame with their
counterparts in the {\sc tycho-2} catalogue \citep{hog2000}and establishes an astrometric solution 
with an RMS precision of 0.1 to 0.2 pixel. 

Aperture photmetry is then carried out at 
the positions on each CCD image of all objects in the USNO-B1.0 catalogue \citep{monet2003} with 
second-epoch red magnitudes brighter than 15.0. Fluxes are measures in three apertures with radii of 
2.5, 3.5 and 4.5 pixels. The ratios between the fluxes in different pairs of apertures 
yield a ``blending index"  which quantifies image morphology, and is used to flag 
blended stellar images and extended, non-stellar objects. 
Individual objects are allocated SuperWASP
 identifiers of the form ``1SWASP J$hhmmss.ss+ddmmss.s$",  which are based on their 
 USNO-B1.0 coordinates for equinox J2000.0 and epoch J2000.0.

The resulting fluxes are corrected for primary and 
secondary extinction, and the zero-point for each frame is tied to a network of local 
secondary standards in each field, whose magnitudes are derived from WASP fluxes
transformed via a colour equation relating instrumental magnitudes to {\sc tycho-2} $V$
magnitudes. 
 The resulting fluxes are stored in the SuperWASP Data Archive at the 
University of Leicester.
The corresponding transformed magnitudes for all objects in each field are referred to as 
 ``WASP $V$" magnitudes throughout this paper.

\subsection{Sample selection and survey completeness}

The field chosen for development and testing of the transit search algorithm is centred at 
$\alpha_{2000} = 01$h 43m, $\delta_{2000}=+31^\circ$ 26'. A search of the
WASP archive, centred on this position and covering the full 7.8$^\circ$-square
field of view of the camera, yielded
light-curves of 7840 stars  brighter  than WASP $V = 13$ for which a catalogue 
query indicated that the light-curves
comprised more than 500  valid photometric data points. Indeed most of the 
objects in this field had more than 2000 valid photometric measurements. 
The maxipara 2 mum number of valid observations in any light-curve was 2301.
The resulting  set of light-curves was loaded into a rectangular matrix of 7840 
light-curves by 2301 observations for further processing.

\begin{figure}
\psfig{figure=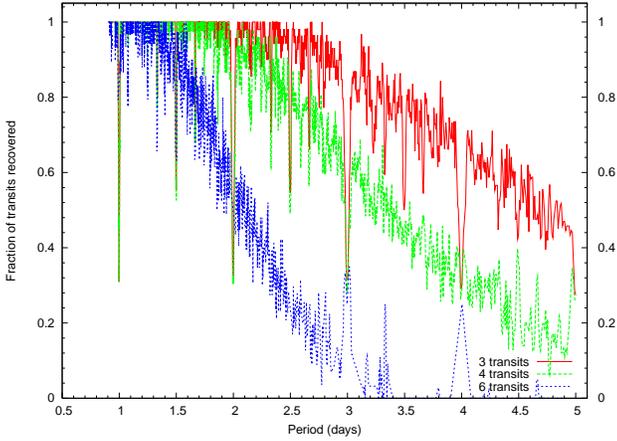,width=8.5cm,angle=270}
\caption[]{Probability of observing more than $N_t$ transit events in the 2004 data from the
field centred at RA = 01h 43m, Dec = +31$^\circ$ 26$'$, as a function of orbital period. For
periods less than 5 days, the probability of observing 3 or more transits is at least 50 percent,
except at periods close to integer and half-integer numbers of days.}
\label{fig:ntrans}
\end{figure}

The expected number of transits present in the observed light-curve of any given star
depends on the sampling pattern of the observations and the period and phase of the
transit cycle. In Fig.~\ref{fig:ntrans} we plot the probability of  $N_t$ or more transits being present in the data, as a function of orbital period.  We consider a transit as having been ``observed'' if data have been obtained within the phase ranges $\phi < 0.1 w/P$ or $\phi > 1-0.1w/p$, where $w$ is the expected transit duration as described at the start of  Section~\ref{sect:algorithms} and $P$ is the orbital period. The regular sampling pattern on most nights of acceptable quality generally ensures that at least 40 percent of a given transit event must be well-observed if it is to be counted according to this criterion. For the field studied here, the prospects of observing at least three transits is 70\%\ or better at periods less than 3.5 days.

\section{Systematic error removal}

The reduced data from the pipeline inevitably contain low-level systematic errors. 
In this section
we describe a coarse initial decorrelation and application of the {\sc SysRem} 
algorithm of \citet{tamuz2005}.
Before modelling and removing patterns of correlated error we perform a coarse 
initial decorrelation by referencing each star's magnitude to its own mean,
removing small night-to-night and frame-to-frame differences in the zero-point, 
and measuring the additional, independent variance components introduced in 
individual stars by their intrinsic variablility and in some observations by patchy cloud.

\subsection{Coarse decorrelation}

We start with a two-dimensional array $m_{ij}$ of processed stellar magnitudes from the 
pipeline. The first index $i$ denotes a single CCD frame within the entire season's data. The 
second index labels an individual star. We compute the mean magnitude of each star:
\begin{equation}
\hat{m}_j=\frac{\sum_i m_{ij} w_{ij}}{\sum_i w_{ij}}
\label{eq:mhat}
\end{equation}
where the weights $w_{ij}$ incorporate both the formal variance $\sigma^2_{ij}$
calculated by
the pipeline from the stellar and sky-background fluxes, and an additional systematic 
variance component $\sigma^2_{t(i)}$ 
introduced in individual frames by passing wisps of cloud, Sahara dust events and 
other transient phenomena which degrade the extinction correction:
\begin{equation}
w_{ij} = \frac{1}{\sigma^2_{ij}+\sigma^2_{t(i)}}.
\label{eq:cloudwts}
\end{equation}
Data points from frames of dubious quality are thus down-weighted. 
The weight is set to zero for any data point flagged by the pipeline as either missing or bad.

The zero-point correction for each frame $i$ follows:
\begin{equation}
\hat{z}_i=\frac{\sum_j (m_{ij}-\hat{m}_j) u_{ij}}{\sum_j  u_{ij}}
\label{eq:zhat}
\end{equation}
In this case the weights are defined as
\begin{equation}
u_{ij}=\frac{1}{\sigma^2_{ij}+\sigma^2_{s(j)}},
\label{eq:varwts}
\end{equation}
where $\sigma^2_{s(j)}$ is an additional variance caused by intrinsic stellar variability. This 
down-weights variable stars in the calculation of the zero-point offset 
for each frame.

Initially we set $\sigma^2_{t(i)}=\sigma^2_{s(j)}=0$, and compute the average magnitude $\hat{m}_j$ for every star $j$ and the zero-point offset $\hat{z}_i$ for every frame. 

To determine the additional variance of $\sigma^2_{s(j)}$ for the intrinsic variability
of a given star $j$ from the data
themselves, we use a maximum-likelihood approach. We define a data vector
$\vec{X}=\{m_{ij},i=1...n\}$ 
containing the light-curve of star $j$,
and a model 
$\vec{\mu}=\{\hat{m}_j+\hat{z}_i,i=1...n\}$.
If both sets of errors are gaussian, then the probability of the $i$th individual observation is
$$
P(\vec{X}_i | \vec{\mu}_i)=\frac{1}{\sqrt{2\pi}\sqrt{\sigma^2_{ij}+\sigma^2_{t(i)}+\sigma^2_{s(j)}}}
\exp\left(-\frac{(m_{ij}-\hat{m}_j-\hat{z}_i)^2}{2(\sigma^2_{ij}+\sigma^2_{t(i)}+\sigma^2_{s(j)})}
\right).
$$ 

The likelihood of the entire data vector for star $j$ given the model is 
$$
L(\vec{\mu})=(2\pi)^{-n/2}\prod_i \left(\frac{1}{\sqrt{\sigma^2_{ij}+\sigma^2_{t(i)}+\sigma^2_{s(j)}}}\right)\exp\left({-\frac{1}{2}\chi^2}\right).
$$
where 
$$
\chi^2=\sum_i\frac{(m_{ij}-\hat{m}_j-\hat{z}_i)^2}{\sigma^2_{ij}+\sigma^2_{t(i)}+\sigma^2_{s(j)}}.
$$

Taking logs and differentiating, we find that the maximum-likelihood value of
$\sigma_{s(j)}$ satisfies
\begin{eqnarray*}
&&\frac{d}{d\sigma_{s(j)}}\left( \chi^2 +  \sum_i \ln(\sigma^2_{ij}+\sigma^2_{t(i)}+\sigma^2_{s(j)}) \right)\\
&&=2\sigma_{s(j)}\left[ \sum_i \frac{1}{\sigma^2_{ij}+\sigma^2_{t(i)}+\sigma^2_{s(j)}}- \sum_i\frac{(m_{ij}-\hat{m}_j-\hat{z}_i)^2}{(\sigma^2_{ij}+\sigma^2_{t(i)}+\sigma^2_{s(j)})^2}\right]
=0.
\end{eqnarray*}
We solve the equation
\begin{equation}
 \sum_i\frac{1}{\sigma^2_{ij}+\sigma^2_{t(i)}+\sigma^2_{s(j)}}- \sum_i\frac{(m_{ij}-\hat{m}_j-\hat{z}_i)^2}{(\sigma^2_{ij}+\sigma^2_{t(i)}+\sigma^2_{s(j)})^2}
=0
\label{eq:sigma_s}
\end{equation}
iteratively for each $\sigma^2_{s(j)}$, holding the $\sigma^2_{t(i)}$ fixed.

We then perform an analogous calculation, summing over all the stars in the $i$th frame and solving 
\begin{equation}
 \sum_j\frac{1}{\sigma^2_{ij}+\sigma^2_{t(i)}+\sigma^2_{s(j)}}- \sum_j\frac{(m_{ij}-\hat{m}_j-\hat{z}_i)^2}{(\sigma^2_{ij}+\sigma^2_{t(i)}+\sigma^2_{s(j)})^2}
=0
\label{eq:sigma_t}
\end{equation}
for each $\sigma^2_{t(i)}$ holding the $\sigma^2_{s(j)}$ fixed.

At this stage we refine the mean magnitude per star, the zero point offsets and the 
additional variances for stellar variability and patchy cloud, by iterating 
Eqs.~(\ref{eq:mhat}), (\ref{eq:zhat}), (\ref{eq:sigma_s}) and (\ref{eq:sigma_t}) to convergence. 

The coarsely decorrelated differential magnitude of each star is then given by 
$$
x_{ij}=m_{ij}-\hat{m}_j-\hat{z}_i.
$$

\subsection{Further decorrelation with {\sc SysRem}}
\label{sect:sysrem}

Some of the many sources of systematic error that affect ultra-wide-field photometry
with commercial camera lenses are readily understood and easily corrected, while
others are less easy to quantify. For example,
the SuperWASP bandpass spans the visible spectrum, introducing 
significant colour-dependent terms into the extinction correction. The pipeline 
attempts to calibrate and remove secondary extinction using {\sc tycho-2} $B-V$ 
colours for the brighter stars, but uncertainties in the colours of the {\sc tycho-2} stars 
and the lack of colour information for the fainter stars means that some systematic 
errors remain. Bright moonlight or reduced transparency arising from stratospheric 
Sahara dust events reduce the contrast between faint stars and the sky background,
altering the rejection threshold for faint sources in the sky-background annulus and biassing
the photometry for faint stars. The SuperWASP camera lenses are vignetted across the
entire field of view, and the camera array is not autoguided, so polar-axis misalignment 
causes stellar images to drift by a few tens of pixels across the CCD each night. It is possible that
temperature changes during the night could affect the camera focus, changing the
shape of the point-spread function across the field and biassing the photometry for 
fainter stars.

\begin{figure}
\psfig{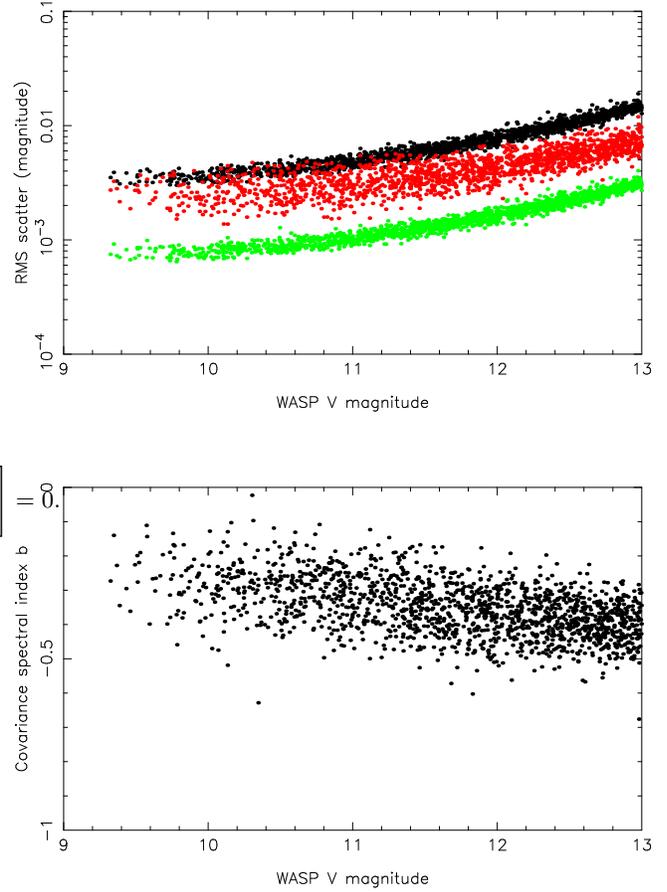}
\caption[]{Upper panel: RMS scatter versus magnitude prior to decorrelation with {\sc SysRem}. The upper curve
shows the RMS scatter of individual data points in the light-curves of the non-variable stars
in the ensemble. The middle curve shows the scatter in the same light-curves after performing 
a moving weighted average over all complete 2.5-hour intervals within each night. The lower curve
shows the RMS scatter of the individual data points divided by the square root of the average 
number of points (typically 22) in a 2.5-hour interval. The correlated noise amplitude among the 
brightest stars is typically 0.0025 mag. Lower panel: Covariance spectral index $b$ as a function of $V$ magnitude prior to decorrelation with {\sc SysRem}. Pure uncorrelated (white) noise should give $b=-0.5$, while pure correlated noise should give $b=0$. We see that the effects of correlated noise are most pronounced for the brightest stars. Even the faintest stars are affected to some extent.}
\label{fig:rmsraw}
\end{figure}

\begin{figure}
\psfig{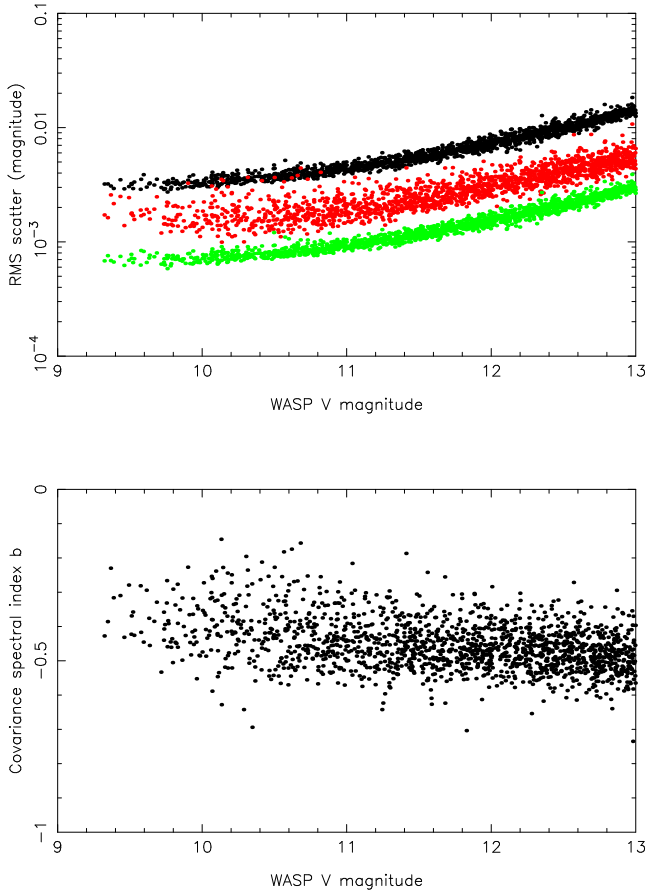}
\caption[]{Upper panel: RMS scatter versus magnitude after removal of the four strongest correlated error components using  {\sc SysRem}. The curves are defined as in Fig.~\ref{fig:rmsraw}. The correlated noise amplitude  of the binned data is reduced to 0.0015 magnitudes for  the brightest stars. Lower panel: Covariance spectral index $b$ as a function of $V$ magnitude after removal of 4 correlated error components using {\sc SysRem}. While some effects of correlated noise remain for the brightest stars, stars fainter than $V=11.0$ have covariance spectral indices close to the value $b=-0.5$ expected for white noise.}
\label{fig:rmstam}
\end{figure}

These systematic errors, and no doubt others as-yet unidentified, have a serious impact on the detection threshold for transits. \citet{pont2006} 
discussed methods of characterising the structure of the covariance matrix for a 
given stellar light-curve. The first and simplest method is to carry out boxcar smoothing
of each night's data, with a smoothing length comparable to the typical 2.5-hour duration 
of a planetary transit. For every set of $L$ points spanning a complete 2.5-hour interval interval starting at the $k$th observation, 
we construct an optimally-weighted average magnitude
$$
\hat{m} _k= \frac{\sum_{i=k}^{k+L-1} m_{i} w_{i}}{\sum_{i=k}^{k+L-1} w_{i}},
$$
with bad observations down-weighted as above using
$w_{i} = 1/(\sigma^2_{i}+\sigma^2_{t(i)})$.

The RMS scatter $\sigma_{\mbox{binned}}$ in the smoothed light-curve of $\hat{m} _k$ values
is then compared to the RMS scatter $\sigma_{\mbox{unbinned}}$ of the individual data points. 
For uncorrelated noise, we expect $\sigma_{\mbox{binned}}= \sigma_{\mbox{unbinned}}/\sqrt{L}$,
where $L$ is the average number of observations made in a 2.5-hour interval. In 
Fig.~\ref{fig:rmsraw} we plot $\sigma_{\mbox{unbinned}}$, $\sigma_{\mbox{binned}}$ and 
$\sigma_{\mbox{unbinned}}/\sqrt{L}$ as functions of $V$ magnitude. For clarity, we exclude all obviously variable stars having $\sqrt{\sigma^2_{s(j)}}>0.005$ magnitude (as defined in Eq.~\ref{eq:varwts}). The
RMS scatter in the binned data is typically 0.0025 magnitude for the brightest non-variable stars,
far worse than the 0.0008 magnitude that would be achieved if the noise were uncorrelated.

The covariance structure of the correlated noise is quantified by the power-law dependence 
of the RMS scatter on the number of observations used in the boxcar smoothing:
$$
\sigma_{\mbox{binned}}=\sigma_{\mbox{unbinned}} L^b.
$$
For completely uncorrelated noise we expect $b = -1/2$, while for completely correlated noise 
(e.g. from intrinsic large-amplitude stellar variability on timescales longer than the longest 
smoothing length considered but
shorter than the data duration) we expect the RMS scatter to be independent of the number of 
data points, giving $b=0$. We measure $b$ for each star using the incomplete smoothing intervals
at the start and end of the night. We create a set of binned magnitudes obtained for $L= 1, 2, 3, ...$ 
consecutive observations. The RMS scatter in the binned magnitudes for each value of $N$ is then 
plotted as a function of $L$ and a power-law fitted to determine $b$. In Fig.~\ref{fig:rmsraw} we plot 
$b$ as a function of $V$ magnitude, again excluding intrinsic variable stars having  $\sqrt{\sigma^2_{s(j)}}>0.005$ 
magnitude. As expected, we find the effects of correlated noise to be most pronounced for
the brightest non-variable stars. Even at our faint cutoff limit of $V= 13$, however, we do not fully 
recover the uncorrelated noise value $b=-0.5$.

We use the {\sc SysRem} algorithm of \citet{tamuz2005} to identify and remove 
patterns of correlated noise in the data. The reader is referred to that paper for details 
of the implementation. The {\sc SysRem} algorithm produces a corrected magnitude
$\tilde{x}_{i,j}$ for star $j$ at time $i$, given by
$$
\tilde{x}_{i,j}=x_{i,j}-\sum_{k=1}^M\ ^{(k)} c_j \ ^{(k)}a_i.
$$
where $M$ represents the number of basis functions (each representing a distinct pattern of systematic error) removed.
An interesting property of the {\sc SysRem} algorithm is that the inverse variance weighted mean value of  each 
basis function, multiplied by the corresponding stellar coefficient, is so close to zero for all but the most highly 
variable stars that it is not necessary to repeat the coarse decorrelation. The inverse variance weighted mean 
change in the zero-point of each frame is generally less than 0.001 magnitude, again rendering further 
coarse decorrelation unnecessary after the final application of {\sc SysRem}.

 \begin{figure*}
\begin{tabular}{cc}
\psfig{figure=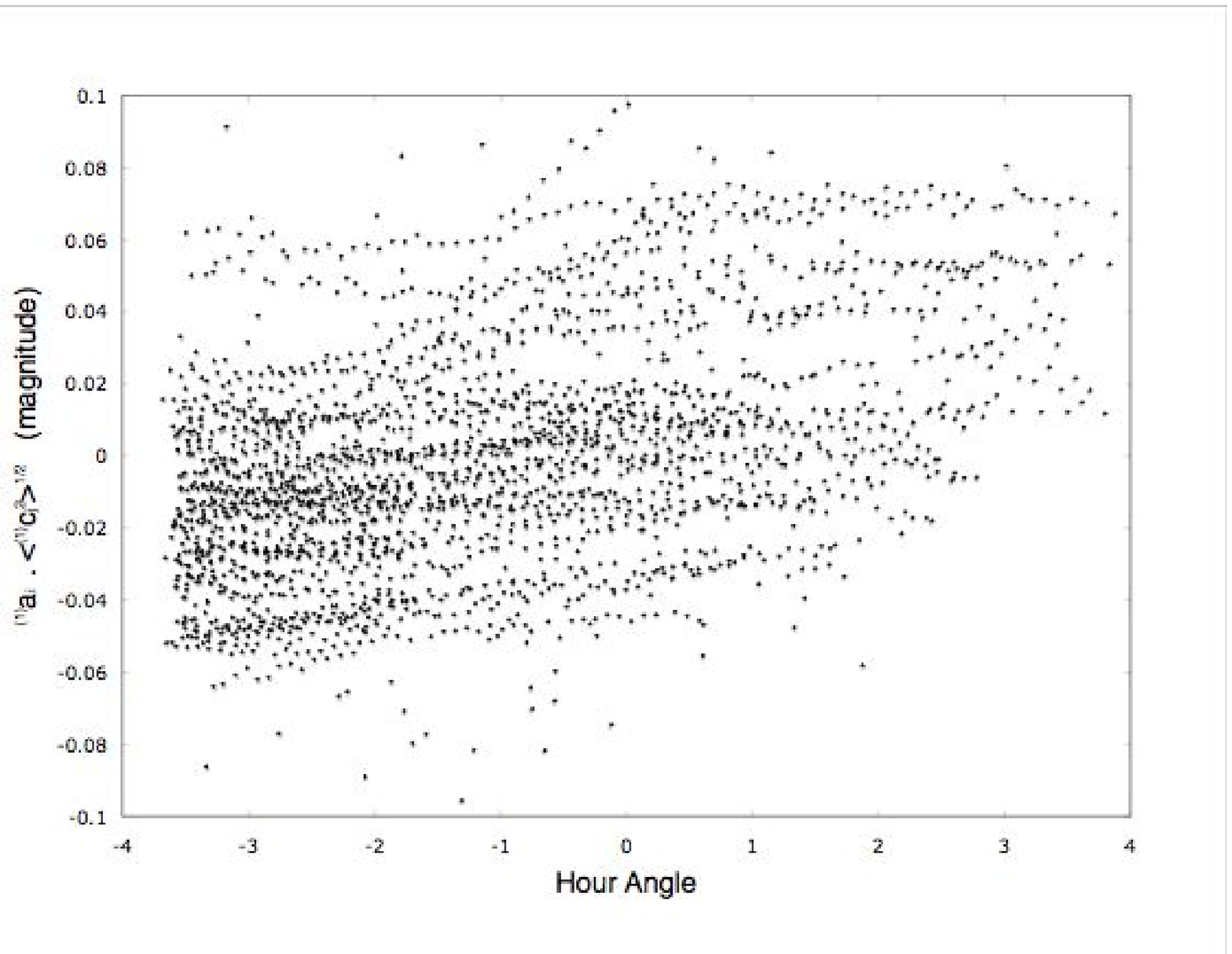,width=8.5cm} &
\psfig{figure=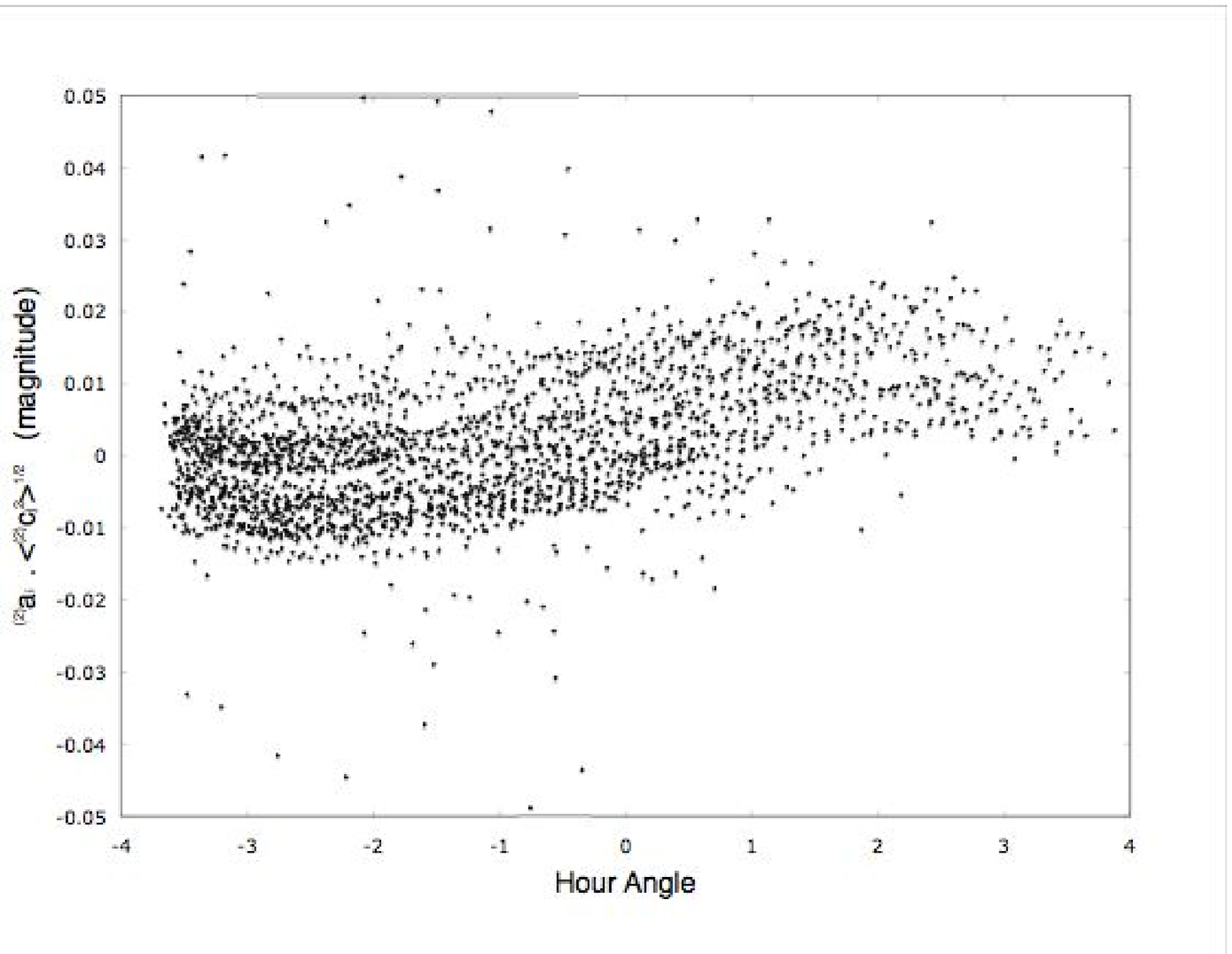,width=8.5cm}\\
\psfig{figure=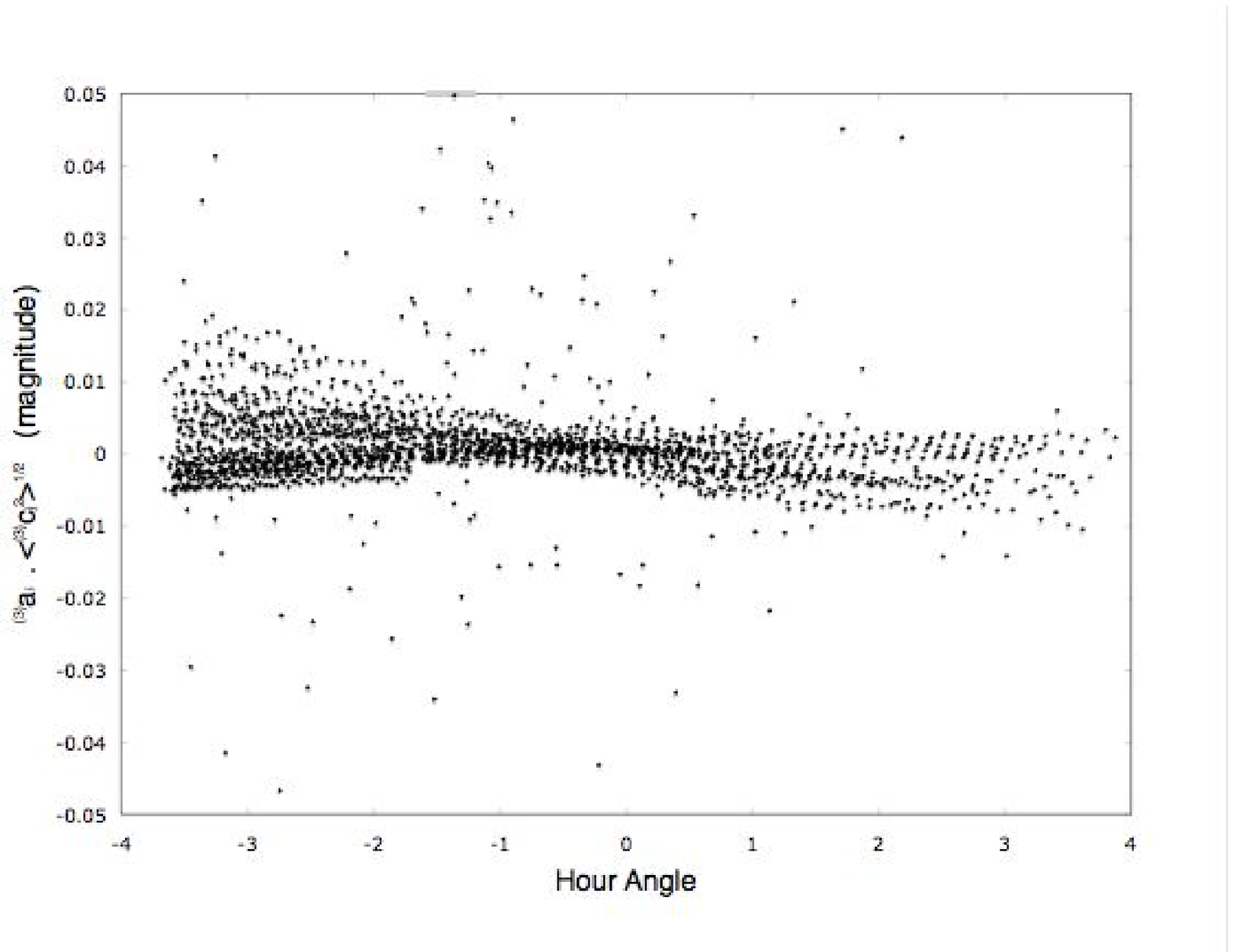,width=8.5cm} &
\psfig{figure=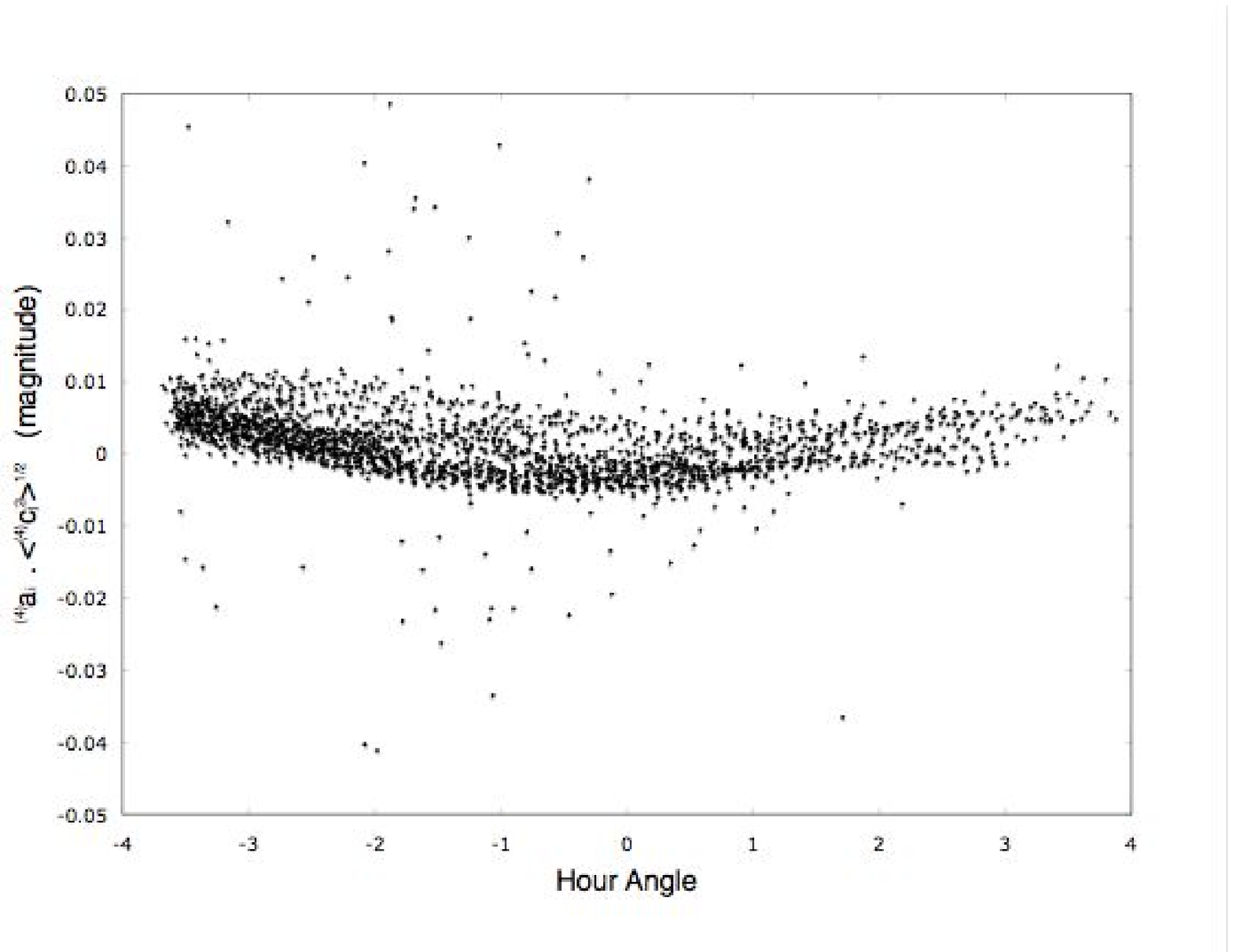,width=8.5cm}
\end{tabular}
\caption[]{The first four {\sc SysRem} basis functions are plotted as a function of hour angle after being folded on a period of 1 sidereal day. To give an idea of the magnitude correction applied to a typical star, each basis function is scaled by the RMS scatter of the corresponding stellar coefficients.}
\label{fig:basfuncs}
\end{figure*}

As described by Tamuz et al, we find several distinct basis functions $a_i$ representing patterns
of correlated systematic error in the data which affect every star in the field to an extent quantified by the 
coefficents $c_j$. In Fig.~\ref{fig:basfuncs} we plot the first four basis functions against hour angle, after folding the entire season's basis-function values on a period of 1 sidereal day.
The first and strongest basis function produced by {\sc SysRem} shows a generally smooth night-to-night variation $^{(1)}a_i$ 
with a characteristic timescale of order 10 days. Superimposed on this large-amplitude, long-timescale 
variation is a small-amplitude, linear trend through each night. Neither variation appears to be related
 directly to either the lunar cycle or to transparency losses caused by intermittent Sahara dust events, 
 but the stellar coefficients $^{(1)}c_j$ show a strong correlation with magnitude at $V>12$. We infer that this 
 systematic error component may be related to a combination of sky brightness and atmospheric 
 transparency that could affect the rejection threshold for faint stellar images in the sky aperture, leading 
 the pipeline to underestimate the brightness of faint stars in bright moonlight and/or poor transparency.

 The second and fourth basis functions resemble half-sinusoids, 90 degrees out of phase, when plotted 
 modulo sidereal time. Both the secondary extinction and the flat-field vignetting correction are expected 
 to vary as functions of sidereal time. As discussed above, SuperWASP is not autoguided, and a small 
 misalignment of the polar axis causes stellar images to re-trace the same path across the CCD, a few 
 tens of pixels long, every sidereal day. The second and fourth basis functions are probably a linear 
 combination of these two effects. The corresponding stellar coefficients $^{(2)}c_j$  and $^{(4)}c_j$
 should be correlated with departures from the stellar colour used for the extinction modelling, and 
 with the gradient of the residuals in the vignetting function along the trajectory of each stellar image. 
 The third basis function is a linear trend through the night, generally increasing but occasionally 
 decreasing. The origin of this component is not obvious, but one possibility is that it could arise from 
 temperature-dependent changes in the camera focus through the night.

The fifth and higher basis functions gave significantly smaller changes in $\chi^2$ than the first four, and 
their form appears to represent mainly stochastic events affecting only a few points in the light-curves 
of a relatively small number of stars. To avoid the danger of removing genuine stellar variability, we 
modelled the global systematic errors using only the first four {\sc SysRem} basis functions.

In Fig.~\ref{fig:rmstam} we show the RMS-magnitude diagram and 
covariance index $b$ as functions of $V$ magnitude after processing with {\sc SysRem}. We find that 
for stars fainter then $V=11.0$ the noise in the corrected light-curves is almost uncorrelated.
For brighter stars some residual evidence of correlated noise remains. On the 2.5-hour timescale of a typical transit, however, the RMS amplitude of the correlated noise component is reduced to 
values of order 0.0015 magnitude with the help of {\sc SysRem}.

\section{Hybrid transit-search algorithms}
\label{sect:algorithms}

We use an adaptation of the BLS algorithm \citep{kovacs2002} for the 
initial search. We set up a coarse search grid of frequencies and transit epochs. 
The frequency step is such that the 
accumulated phase difference between successive frequencies over the full duration 
of the dataset corresponds to the expected width of a transit at the longest period searched. A set of transit epochs is
defined at each frequency, at phase intervals equal to the expected transit width at that frequency. The expected transit duration is computed from the orbital frequency using Kepler's third law assuming a stellar mass of 0.9 M$_\odot$. Since the majority of the main-sequence stars in the magnitude range of interest have masses between 0.7 and 1.3 M$_\odot$ and the transit duration at a given period scales as $(M_*/M_\odot)^{2/3}$, the predicted transit duration is unlikely to be in error by more than 20 to 30 percent even at the extremes of the mass range.

At each trial period and epoch, The transit depth and goodness-of-fit statistic $\chi^2$
are calculated using a variant of the optimal fitting methods of Kovacs et al, reformulated
such that the goodness-of-fit criterion has the dimensions of the $\chi^2$ statistic. A similar approach has also been used by \citet{aigrain2004} and by \citet{burke2006}.

After processing with {\sc SysRem}, the light-curve of a given star comprises  a set of 
observations $\tilde{x}_i$ with associated formal variance estimates $\sigma^2_i$
and additional, independent variances $\sigma^2_{t(i)}$ to account for transient spatial
irregularities in atmospheric extinction. We define inverse-variance weights 
$$
w_i=\frac{1}{\sigma^2_i+\sigma^2_{t(i)}+\sigma^2_{s(j)}},
$$
and subtract the optimal average value 
$$
\hat{x}=\frac{\sum_i \tilde{x}_i w_i}{\sum_i  w_i}
$$
to obtain $x_i=\tilde{x}_i -\hat{x}$. We also define 
$$
t = \sum_i  w_i, 
\hspace{2mm}
\chi^2_0=\sum_i x_i^2 w_i,
$$
summing over the full dataset. Note that the weights defined here include the independent variance component $\sigma^2_{s(j)}$. This has the effect of lowering the significance of high-amplitude variable stars, but has little effect on low-amplitude variables such as planetary transit candidates.

In the BLS method, the transit model is characterised by a periodic box function whose period, 
phase and duration determine the subset $\ell$ of  ``low" points observed while transits are in 
progress. In many implementations the partitioning of the data can be the slowest part of the BLS 
procedure \citep{aigrain2004}. We achieve substantial speed gains by
computing  the orbital phase $\phi$ of each data
point and sorting the phases in ascending order together with their original sequence 
numbers. We partition the phase-ordered data into a contiguous block of out-of-transit
points for which $w/2P<\phi<1-w/2P$, where $w$ is the transit duration and $P$ is the orbital period,
and the complement of this subset comprising the in-transit points. The summations that follow use
the phase-ordered array of sequence numbers to access the in-transit points.

Using notation similar to that of \citet{kovacs2002}, we define
$$
s=\sum_{i\in\ell} x_i w_i,
\hspace{2mm}
r=\sum_{i\in\ell} w_i,
\hspace{2mm}
q=\sum_{i\in\ell} x_i^2 w_i.
$$
The mean light levels inside ($L$) and outside($H$) transit are given by
$$
L=\frac{s}{r},H=\frac{-s}{t-r}
$$
with associated variances
$$
\mbox{Var}(L)=\frac{1}{r},
\hspace{2mm}
\mbox{Var}(H)=\frac{1}{t-r}.
$$
The fitted transit depth and its associated variance are
$$
\delta=L-H=\frac{st}{r(t-r)}, 
\hspace{2mm}
\mbox{Var}(\delta)=\frac{t}{r(t-r)},
$$
so the signal-to-noise ratio of the transit depth is
$$
\mbox{S/N}=s\sqrt{\frac{t}{r(t-r)}}.
$$
The improved fit to the data is given by
$\chi^2=\chi^2_0-\Delta\chi^2$, where the improvement in the fit when 
compared with that of a constant light curve is
$$
\Delta\chi^2=\frac{s^2t}{r(t-r)}.
$$
Note also that $\Delta\chi^2=s\delta=(S/N)^2$.
The goodness of fit to the portions of the light-curve outside transit, where the light 
level should be constant, is
$$
\chi^2_h=\chi^2_0-\frac{s^2}{(t-r)}-q.
$$
The best-fitting model at each frequency is selected, and the corresponding transit 
depth, $\Delta\chi^2$ and $\chi^2_h$ are stored for each star. 

\subsection{Selection of potential candidates}

Following an initial, coarse application of the BLS algorithm, we filter the candidates 
by rejecting obviously variable stars for which the post-fit $\chi^2 > 3.5N$, where N 
is the number of observations. We reject stars for which the best solution has fewer than 
two transits. We also reject those best-fit solutions for which the phase-folded light curve 
contains gaps greater than 2.5 times the expected transit duration. Such solutions  
arise in a small number of stars that suffer from errors in the vignetting correction 
near the extremities of the field of view, or from transient dust motes in the optics that are 
not completely flat-fielded out. Stellar images drift across the image by a few dozen pixels 
during a typical night, because SuperWASP is unguided and has a small misalignment of 
the polar axis. The drift pattern in the light curve
recurs on a period of one sidereal day and its form 
depends strongly on location. It  is significant mainly around the edges of the chip and 
near transient dust-ring features in the flat field. The SysRem algorithm is not effective at 
removing this type of variation from the small number of stars affected, so we reject them
at this stage in the analysis. There is sufficient overlap between cameras that there is a high 
probability of the same star being recorded in a less problematic part of an adjacent 
camera's field of view.

We select candidates for finer analysis by making cuts on  two light-curve statistics, in 
both of which we expect stars showing periodic transit signals to lie well out in one tail 
of  the distribution.  The first is the ``signal-to-red noise" ratio $S_{\mbox{red}}$ of the 
best-fit transit depth to the RMS scatter binned on the expected transit duration:
$$
S_{\mbox{red}} = \frac{\delta\sqrt{N_t}}{\sigma L^b}.
$$
As in Sect.~\ref{sect:sysrem} above, $L$ is the average number of data points spanning a single transit, $b$ is the power-law index that quantifies the covariance structure of the correlated noise, $N_t$ is the number of transits observed, 
$\delta$ is the transit depth and $\sigma$ is the weighted RMS 
scatter of the unbinned data. Since the transit  depth $\delta$ is a signed quantity, 
this statistic distinguishes periodic 
dimmings from periodic brightenings. The second statistic is the ``anti-transit ratio"
$\Delta\chi^2/\Delta\chi^2_-$ proposed by \citet{burke2006}, being the ratio of the strongest 
peak in the periodogram of $\delta\chi^2$ that corresponds to a dimming, 
to the strongest peak corresponding to a brightening. We adopt
conservative thresholds, requiring $S_{\mbox{red}}<-5$ and $\Delta\chi^2/\Delta\chi^2_- > 1.5$. Note
that Burke et al use a threshold of 2.75 for final selection on the latter statistic.

Even this relatively loose set of selection criteria eliminates 95.5 to 97.5 percent of all the 
stars in the sample, typically leaving 100 to 200 surviving objects worthy of more detailed 
study from an initial sample of several thousand stars.

In order to ensure
that the most obvious transit candidates are not rejected by the filtering, we injected 
synthetic patterns of transits, with randomly-generated periods and 
epochs, into the light-curves of 100 randomly-chosen stars in the test dataset. The transits were given 
depths of 0.02 magnitude, and their durations were again computed from the orbital 
frequency using Kepler's third law assuming a stellar mass of 0.9 M$_\odot$. The synthetic transit signatures were added to the actual data, thus preserving the noise properties of the observations.

\begin{figure}
\psfig{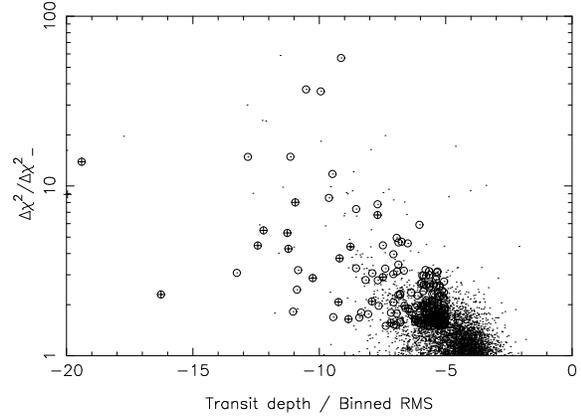}
\caption[]{A scatterplot of the anti-transit ratio against the signal to red noise ratio for stars in the
0143+3126 field shows that the majority of non-variable stars yield anti-transit ratios less than 2.0, 
and spurious best-fit signal to red noise ratios between -3.0 and -6.0. Crosses denote the 100 stars 
for which  synthetic transits with depths of 0.02 magnitude were injected. Stars that satisfy the initial 
selection criteria for refined analysis are circled.}
\label{fig:ant_snr}
\end{figure}

In Fig.~\ref{fig:ant_snr} we plot the anti-transit ratio against  $S_{\mbox{red}}$  for all stars in the dataset with 
positive values of $S_{\mbox{red}}$. The 100 stars for
which synthetic transits were injected are denoted by crosses, and the remainder by dots.
Those stars selected for further study are circled, confirming that the preselection procedure
captures a set of objects that includes nearly all candidates with significant transit signals. Those
that were not selected were either too faint and noisy to yield a significant detection, or were 
superimposed on light-curves of intrinsically variable stars that failed to satisfy the other selection
criteria.

\subsection{Refinement of transit parameters}

The reduced sample is again subjected to a BLS search, this time utilising a finer grid spacing
in which the frequency step is chosen to give a phase drift over the entire data train that is no 
more than half the expected transit width. The grid spacing in epoch is set at half the transit width.
For each star in the sample we identify the five most significant peaks that correspond to 
transit-like dimmings, and the three most significant peaks that correspond to brightenings.  
(Selecting the three most significant peaks at the resolution of the grid search suffices to capture the strongest peak reliably after Newton-Raphson refinement, but for dimmings we are interested in 
possible aliases, so we examine the five most significant peaks).

Having identified a subset of objects in which statistically-significant transit-like signals 
may be present, we next refine the transit parameters. Instead of using a pure box function we 
adopt a softened box-like function $\mu(t_i)$ developed by \citet{protopapas2005}:
$$
\mu(t_i)=\frac{1}{2}\delta
\left(
\tanh
\left[
c(t_p+\frac{1}{2})
\right]
+
\tanh
\left[
c(t_p-\frac{1}{2})
\right]
\right)
$$
to approximate the light-curve at time $t_i$, where 
$$
t_p=\frac{P\sin[\pi (t-T_0)/P]}{\pi \eta}.
$$ 
Here $T_0$ is the epoch of mid-transit, $P$ is the orbital period, $\delta$ is the 
transit depth, $\eta$ is the transit duration and $c$ is a softening parameter that 
determines the duration of ingress and egress in the model transit.

This function has the advantage that it is analytically differentiable with respect to 
the key transit parameters $T_0$, $P$, $w$ and $\delta$, which means that we can 
refine them quickly and efficiently using a Newton-Raphson approach based on 
the derivative functions $d\mu/dT_0$, $d\mu/dP$ and $d\mu/d\eta$. 

For an estimated set of parameters $T_0$, $P$, $\eta$ we fit the transit depth by defining
a basis function
$$
p_i = d\mu(t_i)/d\delta =\mu(t_i)/\delta
$$ 
and, for each observation $x_i$, determine the optimal scaling factor to fit the 
light-curve:
\begin{equation}
\hat{\delta}=\frac{\sum_i (x_i-\hat{x})(p_i-\hat{p})w_i}{\sum_i (p_i-\hat{p})^2 w_i}
\label{eq:deltahat}
\end{equation}
where $\hat{x}=\sum_i x_i w_i/\sum_i w_i$ and
$\hat{p}=\sum_i p_i w_i/\sum_i w_i$. 

At this stage in the analysis we omit the variance component $\sigma^2_{s(j)}$
due to stellar variability,  but retain the patchy-extinction contribution $\sigma^2_{t(i)}$
so that poor quality data are down-weighted correctly:
$$
w_i=\frac{1}{\sigma^2_i+\sigma^2_{t(i)}}.
$$

We refine the model parameters $T_0$, $P$ and $\eta$ in turn. 
To refine the epoch of transit, for instance, we 
subtract the current model from the data to obtain 
$$
y_i = x_i -\mu(t_i),
$$
and define the basis function 
$$
q_i=d\mu(t_i)/dT_0.
$$
We then determine the optimal scaling factor
\begin{equation}
\hat{dT}_0 = \frac{\sum_i (y_i-\hat{y})(q_i-\hat{q})w_i}{\sum_i (q_i-\hat{q})^2 w_i}
\label{eq:dt0hat}
\end{equation}
where $\hat{y}=\sum_i y_i w_i/\sum_i w_i$ and
$\hat{q}=\sum_i q_i w_i/\sum_i w_i$. 

The new estimate of the transit epoch is $T_0+dT_0$,
so we re-compute $\mu$ and redetermine the transit depth using Eq.~\ref{eq:deltahat}.
The period $P$ and transit width $\eta$ are refined in turn using procedures exactly 
analogous to Eq.~\ref{eq:dt0hat}. The parameter values converge rapidly to the optimal 
solution after a few iterations.

At this stage the phased light-curves of the best-fit solutions are inspected visually to
pick out those candidates showing clear transit-like signatures. Candidates
identified in this way from the 0143+3126 field  are listed in 
Table~\ref{tab:filters1}, and a representative selection of their phased light-curves
are shown in Fig.~\ref{fig:lightcurves}. 
Some of these are
clearly eclipsing binaries, exhibiting secondary eclipses, out-of-transit variability, or
both. Further physical characterisation is needed to distinguish plausible planetary
transit candidates from probable stellar impostors.

\begin{figure*}
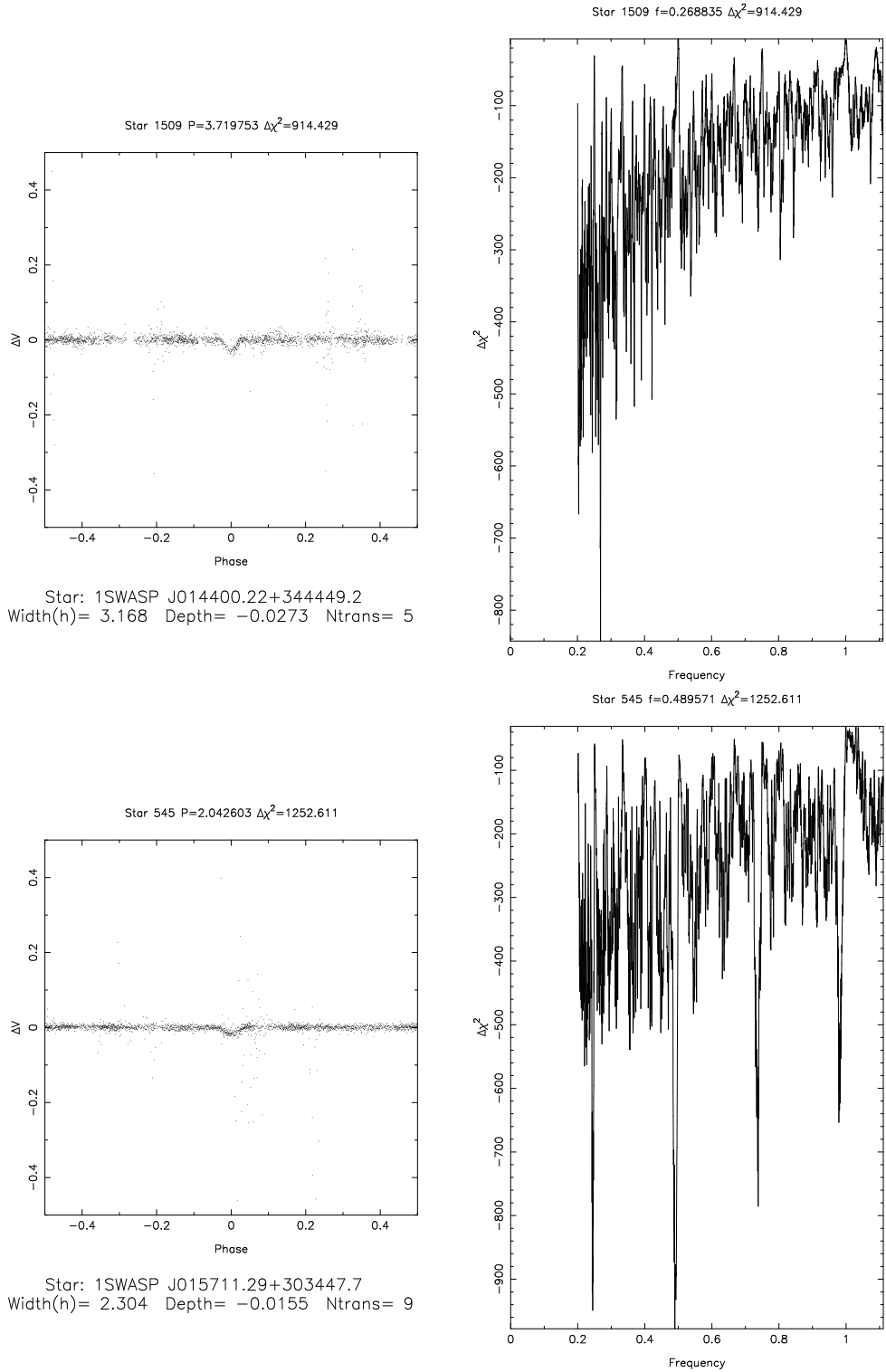

\begin{tabular}{c}
\psfig{figure=J014400.ps,height=10cm,angle=270}\\
\psfig{figure=J015711.ps,height=10cm,angle=270}\\
\end{tabular}
\caption[]{Representative light curves and periodograms
of four objects in the 0143+3126 field exhibiting
transit-like behaviour. The fourth object, 
1SWASP J014228.76+335433.9, shows clear ellipsoidal variability 
outside transit, indicating a stellar binary.}
\label{fig:lightcurves}
\end{figure*}

\begin{figure*}
\begin{tabular}{c}
\psfig{figure=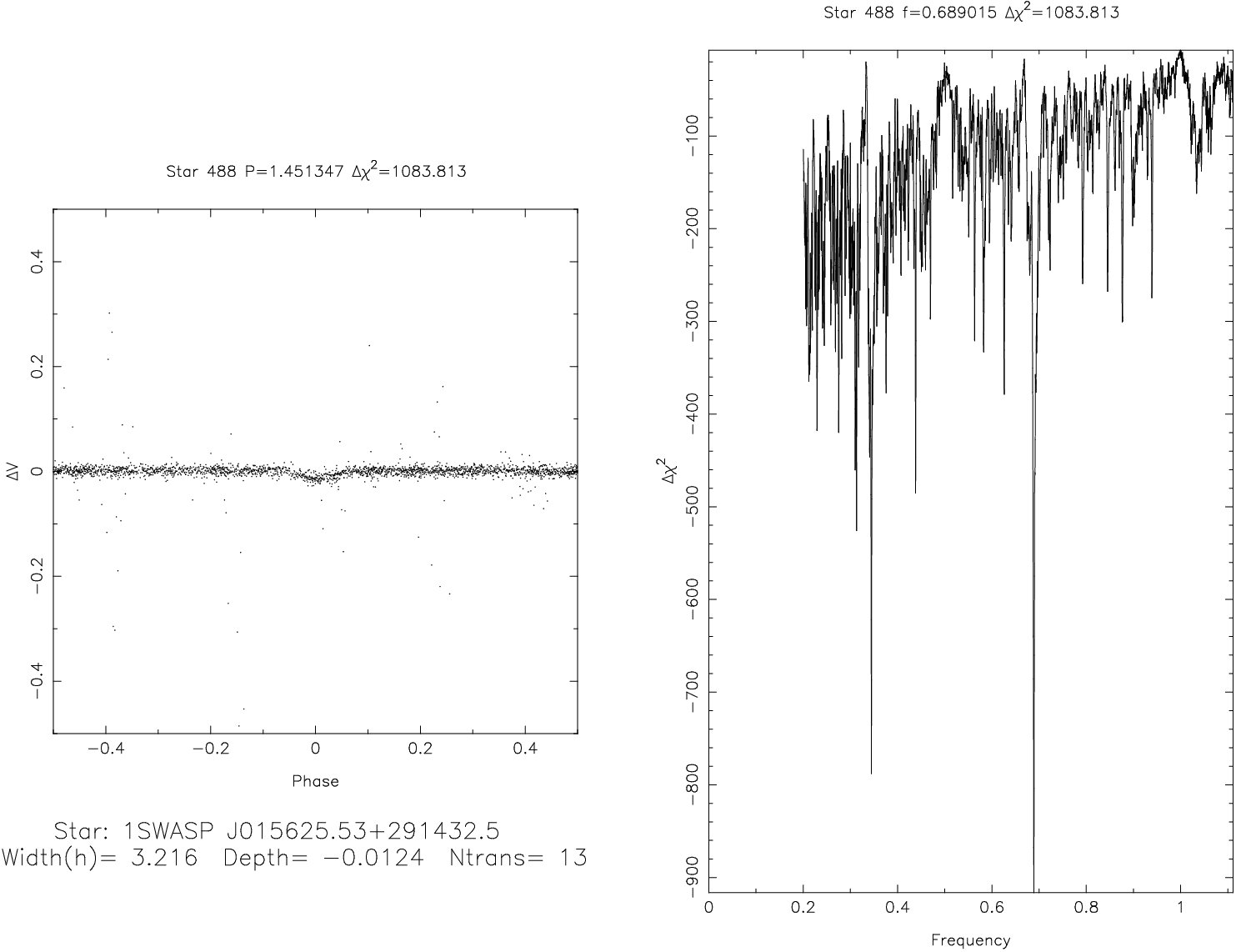,height=10cm,angle=270}\\
\psfig{figure=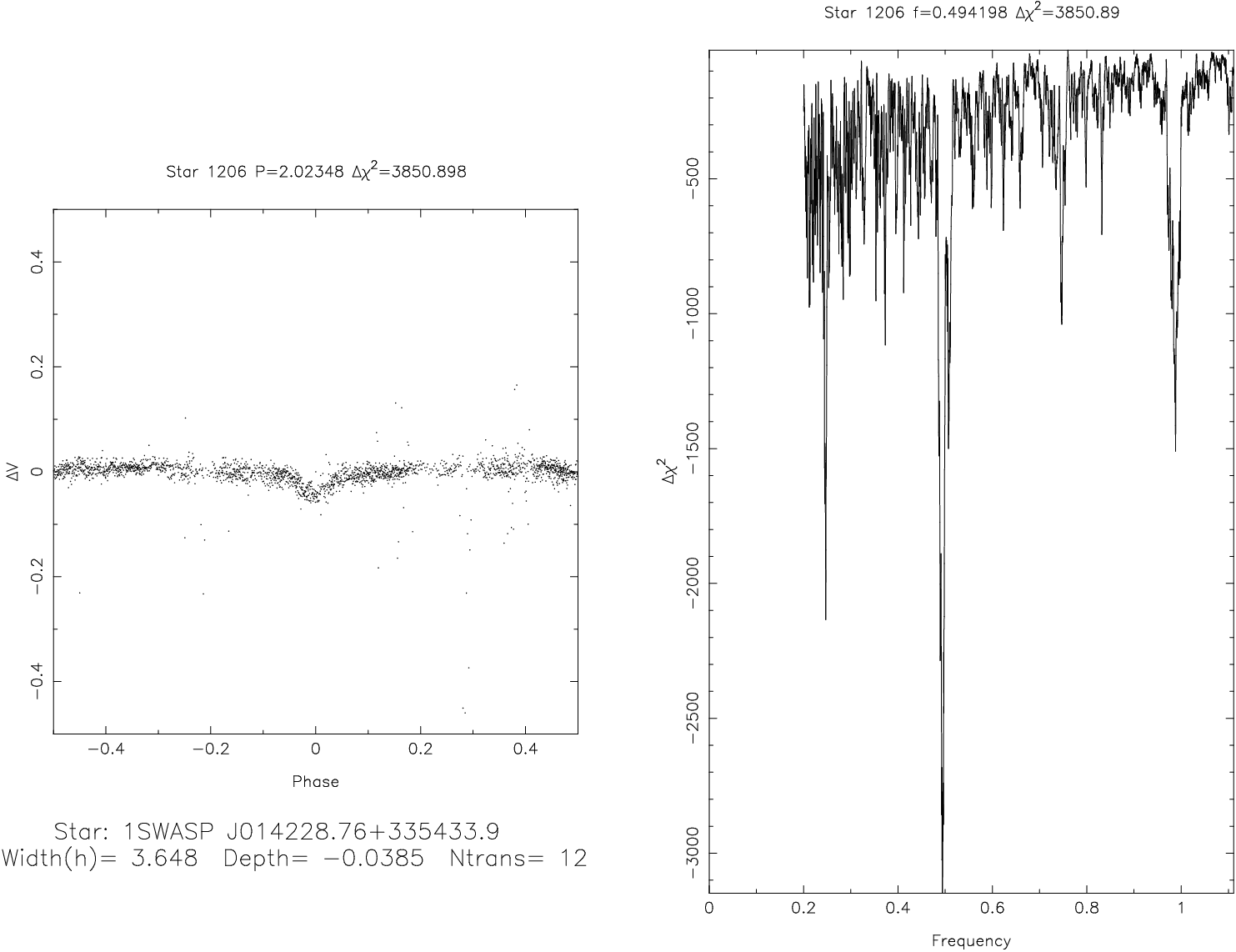,height=10cm,angle=270}\\
\end{tabular}
\contcaption{ }
\end{figure*}

\section{Characterization of candidates}

Both theory \citep{brown2003} 
and the experience of previous transit follow-up campaigns
(\citealt{alonso2004};  
\citealt{bouchy2005}; 
\citealt{pont2005}; 
\citealt{odonovan2006})
indicate that among our transit candidates,
stellar binaries will outnumber genuine planetary transits by an order of magnitude.
Some are grazing eclipsing binaries; others are multiple systems in which the light
of a stellar eclipsing pair is diluted; others still have low-mass stellar or
brown-dwarf companions whose radii are similar to those of gas-giant planets. In
order to mitigate the false-alarm rate our candidates must pass a number of tests
before being considered as high-priority spectroscopic targets. 

\subsection{Ellipsoidal variations}

\citet{drake2003} and \citet{sirko2003}  pointed out that ellipsoidal variables can be 
rejected with a high degree of certainty as stellar binaries. Detached stellar binaries with 
orbital periods of 1 to 5 days can easily be mistaken for 
transiting exoplanet systems if they exhibit either grazing eclipses or deeper eclipses
diluted by the light of a blended third star. In either case, the equipotential surfaces
of the two stars will often be sufficiently ellipsoidal to yield detectable out-of-transit variability.

Since the data have already been partitioned  into a subsets of points inside and
outside transit, we approximate the ellipsoidal variation with phase angle $\theta_i$
as
$$
p_i=-\cos 2\theta_i 
$$
to the out-of-transit residuals $x_i-H$. We sum over the subset $h$ of points outside 
transit to define
$$
u=\sum_h (x_i-H)p_i w_i,\hspace{1cm}v=\sum_h p_i^2 w_i
$$
and so obtain both the amplitude and formal variance 
$$
\epsilon=\frac {u}{v},\hspace{1cm} \mbox{Var}(\epsilon)=\frac{1}{v}
$$
of the ellipsoidal variation. The signal-to-noise ratio of the amplitude is
$$
\mbox{S/N}=\sqrt{\Delta\chi^2_\epsilon}=\frac{u}{\sqrt{v}}.
$$
Any candidate for which  the sign of $\epsilon$ indicates that the system is significantly 
brighter at quadrature than at conjunction, with S/N$ >5$ or so, is noted as a probable stellar 
impostor. In Table~\ref{tab:filters1} we see that two of the possible transit-like candidates 
identified by eye are disqualified in this way. 

The 5$\sigma$ detection threshold for ellipsoidal variation is nearly independent of orbital 
period, but ranges from to 0.003 magnitude at $V=12.5$ to 0.001 mag at $V=10$ for the 
dataset studied here. The expected amplitude of ellipsoidal variation depends on the ratio 
of the primary radius to the orbital separation, and on the mass ratio of the system. 

The SuperWASP transit search yields many objects in which a solar-type star appears in a 
1 to 2-day orbit about a companion with a radius of 1 to 2 R$_{\mbox{Jup}}$. Simple 
projected-area calculations based on a standard Roche equipotential surface model for a 
main-sequence star of 1 M$_\odot$ with a 0.2 M$_\odot$ companion in a 1.5-day orbit 
yield an ellipsoidal variation of order 0.002 magnitude, which should be detectable with 
high significance in an isolated system with $V<11$ or so. Companions less massive than 
this may not yield detectable ellipsoidal variations, but are sufficiently interesting in their 
own right to be worth following up.

Ellipsoidal variability can also help to eliminate impostor systems where a bright foreground 
star is blended with a background eclipsing binary. If one component of an eclipsing 
stellar binary is evolved, such as in an RS CVn system with a 2.5  to 3.0 R$_\odot$ 
K subgiant and a solar-type main-sequence star in a 3-day orbit, the ellipsoidal 
variation can be as great as 0.02 to 0.05 magnitude. If such a system exhibits a 
partial primary eclipse 0.1 to 0.3 magnitude deep and a shallow secondary eclipse, 
and is blended with a foreground star 2 to 3 magnitudes brighter, it can mimic an 
exoplanetary transit. The ellipsoidal variation is thus great enough to remain 
detectable even when diluted by a blend 2 or 3 magnitudes brighter. 

\begin{table*}
\caption{Transit candidates identified by eye from 0143+3126 field.
Quantities that disqualify a candidate from further consideration are shown in boldface. 
IDs of disqualified candidates are shown in parentheses.}
\label{tab:filters1}
\begin{tabular}{cccccccc}
\hline\\
SuperWASP ID	&	$S_{\mbox{red}}$	&	$\delta$	&	Duration (h)	&	Epoch	&	Period	&	$N_t$	&	$(S/N)_{\mbox ellip}$	\\
\hline\\
(1SWASP J015721.21+333517.9)	&	27.4	&	0.0518	&	1.704	&	3182.1328	&	4.14907	&	\bf{2}	&	2.1	\\
1SWASP J013901.75+333640.6	&	21.2	&	0.1781	&	2.04	&	3180.9563	&	3.440225	&	5	&	3.6	\\
1SWASP J015711.29+303447.7	&	12.3	&	0.0155	&	2.304	&	3182.5613	&	2.042603	&	9	&	1.3	\\
(1SWASP J014228.76+335433.9)	&	13.4	&	0.0385	&	3.648	&	3180.7258	&	2.02348	&	12	&	\bf{16.5}	\\
1SWASP J014400.22+344449.2	&	11.9	&	0.0273	&	3.168	&	3180.2839	&	3.719753	&	5	&	4.4	\\
1SWASP J015625.53+291432.5	&	12.2	&	0.0124	&	3.216	&	3182.5415	&	1.451347	&	13	&	4.9	\\
1SWASP J014212.56+341534.4	&	11.3	&	0.0632	&	2.904	&	3182.0596	&	4.305729	&	6	&	4.7	\\
1SWASP J014211.84+341606.5	&	11.1	&	0.0544	&	3.072	&	3182.0503	&	4.30604	&	6	&	4.2	\\
(1SWASP J012536.11+341423.8)	&	8.9	&	0.0623	&	2.52	&	3181.6191	&	1.891481	&	4	&	\bf{9.5}	\\
1SWASP J014549.24+350541.9	&	6.9	&	0.0081	&	2.256	&	3182.407	&	1.452465	&	9	&	0.6	\\
1SWASP J014700.48+280243.6	&	6.7	&	0.008	&	4.32	&	3182.1853	&	1.68013	&	13	&	0.5	\\
\hline\\
\end{tabular}
\end{table*}

\subsection{Stellar mass and planet radius}

\citet{seager2003} and \citet{tingley2005} have developed
methods for deriving the physical parameters of transit candidates based on 
light-curve parameters alone. Seager \&\ Mallen-Ornelas use the orbital period,
the total transit duration, the duration of ingress and egress and the transit depth 
to derive the impact parameter, the orbital inclination, the stellar mass, the 
planet radius and the orbital separation. Because the duration of ingress and egress 
are difficult to measure reliably in noisy data, we define a simplified consistency
test predicated on the assumption that the orbital inclination is close to 90 degrees.
Our aim is simply to estimate the mass of the star and the radius of the planet, and
thereby to determine whether the transits could plausibly be caused by a roughly 
Jupiter-sized planet orbiting a star of roughly solar mass and radius.

Once an object is
found to display transit-like events (characterised by a flat light-curve outside
eclipse, no secondary eclipse and a transit depth $< 0.1$ mag) we search the
USNO-B1.0 catalogue for blends less than 5 mag fainter, located within the 48-arcsec
radius of the 3.5-pixel photometric aperture. We estimate a main-sequence radius and
mass from a $V-K$ colour index derived from the SuperWASP $V$ magnitude
and the 2MASS $K$ magnitude. We infer both the expected transit duration and
the radius of the putative planet using the simplified mass-radius relation of 
\citet{tingley2005},
$$
R_\star\simeq M_\star^{4/5}.
$$
The size of the planet follows from the approximate expression of \citet{tingley2005} for the transit depth $\delta$ for a limb-darkened star:
$$
\frac{R_p}{R_\star}\simeq\sqrt{\frac{\delta}{1.3}}.
$$

High-priority candidates must
display multiple transits, a fitted transit duration no more than 1.5 times more or
less than the predicted value, a transit depth indicating a planetary radius less
than 1.6 Jupiter radii, and have no blends less than 3 magnitudes fainter located
within the 48 arcsec photometric aperture. Their proper motions 
(from the {\em Hipparcos},  {\sc tycho-2} or USNO-B1.0 catalogues) and $V-K$ colours
must also be consistent with main-sequence stars rather than giants, the luminosity
class being inferred from the reduced 
proper-motion method of \citet{gould2003} and the 
giant-dwarf separation method of \citet{bilir2006}.

In Table ~\ref{tab:filters2} we list the effective temperatures, spectral types and stellar radii estimated 
from the $V-K$ colour indices, together with the inferred planet radius and the 
ratio $\eta$ of the observed to the expected transit duration. The effective temperatures are 
derived using the calibration of \citet{blackwell1994}, and the radii using the 
interferometrically-determined colour-surface brightness relations of \citet{kervella2004}
together with {\em Hipparcos} parallaxes, where available.
Two objects are rejected
immediately, because brighter stars are found within the radius of the photometric
aperture. Both stars found previously to exhibit significant ellipsoidal variations 
yield inferred companion radii 2.46 and 1.63 R$_{\mbox{Jup}}$, substantially
greater than expected for gas-giant planets. Four of the remaining candidates
are rejected on the same grounds. 

Of the original eleven candidates, three remain. 1SWASP J015625.53+291432.5
and 1SWASP J014549.24+350541.9 are mid-K stars, for which
the inferred companion radii are substantially less than that of Jupiter. Both of these 
have stars less than 5 magnitudes fainter located within the photometric aperture,
so further follow-up is warranted to eliminate the possibility that the blended stars
could be eclipsing binaries. The transit detection in 1SWASP J014549.24+350541.9 is of
rather marginal significance, with $S_{\mbox{red}}=6.93$. The brightest of the three 
candidates, 1SWASP J015711.29+303447.7, appears to be an F6 star with a 
1.38 R$_{\mbox{Jup}}$ companion.

\begin{table*}
\caption[]{Estimated physical parameters for transit candidates in 0143+3126
field. Parameter values inconsistent with planetary status are highlighted in bold type.
IDs of stars disqualified on these grounds are shown in parentheses.}
\label{tab:filters2}
\begin{tabular}{cccccccccc}
\hline\\
SuperWASP ID	&	$V_T$(SW)	&	$V-K$	&	$\rmsub{T}{eff}$	&	Sp. type	&	$R_*/R_\odot$	&$R_p/R_{\mbox Jup}$	&	$\eta$	&	$\rmsub{N}{brighter}$	&	$N_{< 5{\mbox {mag fainter}}}$	\\
\hline\\
(1SWASP J015721.21+333517.9)	&	10.982	&	1.36	&	6118	&	F8	&	1.18	&	\bf{2.29}	&	0.44	&	0	&	0	\\
(1SWASP J013901.75+333640.6)	&	12.986	&	1.86	&	5498	&	G8	&	0.89	&	\bf{3.2}	&	0.58	&	0	&	2	\\
1SWASP J015711.29+303447.7	&	10.352	&	1.19	&	6354	&	F6	&	1.3	&	1.38	&	0.77	&	0	&	0	\\
(1SWASP J014228.76+335433.9)	&	10.963	&	0.93	&	6740	&	F2	&	1.47	&	\bf{2.46}	&	1.08	&	0	&	1	\\
(1SWASP J014400.22+344449.2)	&	11.164	&	1.3	&	6200	&	F8	&	1.22	&	\bf{1.72}	&	0.88	&	0	&	1	\\
1SWASP J015625.53+291432.5	&	10.294	&	2.3	&	5044	&	K3	&	0.76	&	0.72	&	1.67	&	0	&	1	\\
(1SWASP J014212.56+341534.4)	&	12.247	&	1.37	&	6105	&	F9	&	1.17	&	\bf{2.51}	&	0.74	&	0	&	1	\\
(1SWASP J014211.84+341606.5)	&	12.359	&	n/a	&	n/a	&	n/a	&	n/a	&	n/a	&	n/a	&	\bf{2}	&	2	\\
(1SWASP J012536.11+341423.8)	&	12.336	&	2.26	&	5081	&	K2	&	0.77	&	\bf{1.64}	&	1.07	&	0	&	1	\\
1SWASP J014549.24+350541.9	&	11.44	&	2.45	&	4908	&	K4	&	0.73	&	0.56	&	1.22	&	0	&	2	\\
(1SWASP J014700.48+280243.6)	&	11.764	&	n/a	&	n/a	&	n/a	&	n/a	&	n/a	&	n/a	&	\bf{1}	&	1	\\
\hline\\
\end{tabular}
\end{table*}

\section{Discussion and conclusions}

In this paper we have described the methodology that we have adopted for 
searching for transits in the large body of data produced by the SuperWASP 
camera array on La Palma during its first few months of operation, and applied
it to the 7840 stars brighter than $V=13.0$ in a survey field centred at RA  01h 43m,
Dec +31$^\circ$ 26'.

We find that an initial search using the BLS method on a coarse grid of transit 
epochs and orbital frequencies is sufficient for us to eliminate more than 95
percent of the stars searched. This allows us to perform a finer grid search 
on only the remaining 198 stars in the field under consideration. 
We have adapted the analytic Newton-Raphson method of \citet{protopapas2005}
to refine the orbital solutions around periodogram peaks found with the BLS method. 
This method quickly yields the depth and duration of the transits, and the 
frequency and phase of the photometric orbit, while keeping the dimensionality
of the search grid (and hence the processing time required) as low as possible.

We use the ellipsoidal-variation
methodology of \citet{drake2003} and \citet{sirko2003} in our light-curve 
modelling to eliminate probable
stellar binaries. We use the publicly-available 2MASS and USNO-B1.0 
catalogues to obtain colours and proper 
motions for candidates, and to estimate the stellar and planetary radii using methods
similar to those of \citet{seager2003} and \citet{tingley2005}.
These methods confirm that two of our most significant transit detections in this field, and
one more marginal one, have transit properties fully consistent with those expected for 
planets with radii comparable to or somewhat smaller than Jupiter.

The survey field chosen to illustrate the method is only one of more than 100 such
regions surveyed during the course of 2004. Candidates from the other fields will
be presented and discussed in subsequent papers. Together with the three 
candidates presented here, all likely planetary-transit candidates will be subjected to 
high-resolution spectroscopic  follow-up in the latter half of 2006, using
the methodology employed successfully by \citet{bouchy2005} for OGLE-III follow-up.

\section*{Acknowledgments}

The WASP Consortium consists of representatives from the Universities of Cambridge
(Wide Field Astronomy Unit), Keele, Leicester, The Open University, Queens University
Belfast and St Andrews, along with the Isaac Newton Group (La Palma) and the Instituto
de Astrophysic de Canarias (Tenerife). The SuperWASP and WASP-S Cameras were
constructed and operated with funds made available from Consortium Universities and
PPARC. This publication makes use of data products from the Two Micron All Sky Survey, 
which is a joint project of the University of Massachusetts and the Infrared Processing and 
Analysis Center/California Institute of Technology, funded by the National Aeronautics and 
Space Administration and the National Science Foundation.This research has made use of 
the VizieR catalogue access tool, CDS, Strasbourg, France.

\bibliographystyle{mn2e}

\bsp

\label{lastpage}

\end{document}